\documentclass[final,times,12pt]{elsarticle}
\usepackage[a4paper, total={6.5in, 9in}]{geometry}

\usepackage{amssymb}
\usepackage{amsmath}
\usepackage[hidelinks]{hyperref}
\hypersetup{
    colorlinks=true,
    linkcolor=black,
    filecolor=magenta,      
    urlcolor=blue,
}

\usepackage{lineno}

\journal{Journal of Pathology Informatics}

\usepackage[acronym,automake]{glossaries}

\makeglossaries
\renewcommand{\glossaryentrynumbers}[1]{}
\renewcommand{\glossarysection}[2][]{}
\newacronym{wsi}{WSI}{whole slide image}
\newacronym{wsv}{WSV}{whole slide viewer}
\newacronym{idc}{IDC}{imaging data commons}
\newacronym{ife}{IFE}{Iris File Extension}
\newacronym{api}{API}{application programming interface}
\newacronym{tiff}{TIFF}{Tagged Image File Format}
\newacronym{dicom}{DICOM}{Digital Imaging and Communications in Medicine}
\newacronym{ims}{IMS}{image management system}
\newacronym{iod}{IOD}{image object definition}
\newacronym{nema}{NEMA}{National Electrical Manufacturers Association}
\newacronym{nas}{NAS}{network-attached storage}
\newacronym{wasm}{WASM}{WebAssembly module}
\newacronym{ieee}{IEEE}{Institute of Electrical and Electronics Engineers}
\newacronym{dzi}{DZI}{Deep Zoom image}
\newacronym{json}{JSON}{JavaScript Object Notation}
\newacronym{pacs}{PACS}{Picture Archiving and Communication System}
\newacronym{jpeg}{JPEG}{Joint Photographic Experts Group}
\newacronym{avif}{AVIF}{Alliance for Open Media Video 1 file format}
\newacronym{xml}{XML}{Extensible Markup Language}
\newacronym{iec}{IEC}{International Electrotechnical Commission}
\newacronym{i2s}{I2S}{Iris Interoperability Standard}
\newacronym{icc}{ICC}{International Color Consortium}
\newacronym{ascii}{ASCII}{American Standard Code for Information Interchange}
\newacronym{utf8}{UTF-8}{Unicode Transformation Format, 8-bit}
\newacronym{sof}{SOF}{start of file}
\newacronym{eof}{EOF}{end of file}
\newacronym{tpt}{TPT}{time per tile}
\newacronym{fov}{FOV}{field of view}
\newacronym{uid}{UID}{unique identifier}

\begin{document}

\begin{frontmatter}



\title{The Iris File Extension}

\author[MM]{Ryan Erik Landvater MD MEng \fnref{corAuthor}}
\author[MM]{Michael Olp MD}
\author[MM]{Mustafa Yousif MD}
\author[MM]{Ulysses Balis MD}
\affiliation[MM]{organization={University of Michigan Medical School, Department of Pathology},
            addressline={2800 Plymouth Road},
            city={Ann Arbor},
            postcode={48109-2800},
            state={MI},
            country={USA}
}
\fntext[corAuthor]{Corresponding Author}

\begin{abstract}

A modern digital pathology vendor-agnostic binary slide format specifically targeting the unmet need of efficient real-time transfer and display has not yet been established. The growing adoption of digital pathology only intensifies the need for an intermediary digital slide format that emphasizes performance for use between slide servers and image management software. The DICOM standard is a well-established format widely used for the long-term storage of both images and associated critical metadata. However, it was inherently designed for radiology rather than digital pathology, a discipline that imposes a unique set of performance requirements due to high-speed multi-pyramidal rendering within whole slide viewer applications. Here we introduce the Iris file extension, a binary container specification explicitly designed for performance-oriented whole slide image viewer systems. The Iris file extension specification is explicit and straightforward, adding modern compression support, a dynamic structure with fully optional metadata features, computationally trivial deep file validation, corruption recovery capabilities, and slide annotations. In addition to the file specification document, we provide source code to allow for (de)serialization and validation of a binary stream against the standard. We also provide corresponding binary builds with C++, Python, and JavaScript language support. Finally, we provide full encoder and decoder implementation source code, as well as binary builds (part of the separate Iris Codec Community module), with language bindings for C++ and Python, allowing for easy integration with existing WSI solutions. We provide the Iris File Extension specification openly to the community in the form of a Creative Commons Attribution-No Derivative 4.0 International license.

\end{abstract}

\begin{keyword}
Digital Pathology \sep Whole slide image \sep Performance Digital Pathology \sep Iris \sep File format \sep File specification \sep DICOM



\end{keyword}

\end{frontmatter}



\section{Introduction}

Digital pathology, powered by \glspl{wsi}, has the potential to transform diagnostic workflows. Despite this, its widespread adoption has been hampered by technical limitations -- including the challenge of rendering gigapixel images as swiftly and smoothly as viewing glass slides. While advancements in the underlying rendering engines promise improved viewer responsiveness \cite{Landvater2025, Schuffler2022}, the methods we employ to access slide image data locally or remotely must improve simultaneously for these advancements to be realized. 

The introduction of digital slide scanning led to a surge in vendor-specific proprietary \gls{wsi} file formats. Many of these were derived from the \gls{tiff}, likely due to its extensibility and support for random access. The diversity of file structures necessitated the development of open-source translation libraries that provide a standardized \gls{api}, most notably OpenSlide in 2013 \cite{Goode2013} and the Open Microscopy Environment's Bio-Formats. The \gls{dicom} specification, introduced by the \gls{nema} in the 1980s and widely accepted following publication of its current edition in 1993, added support for \gls{wsi} in 2010 \cite{Clunie2021}. This ushered in a vendor-agnostic approach to \gls{wsi} storage, but did not seismically alter the specification metadata structure, tailored for radiology, to adapt to the unique demands of pathology, such as native tile indexing.

While the \gls{dicom} standard is essential for intra- and inter-institutional interoperability, including the archiving of the medical record, the \gls{dicom} specification was not intended to address the unique demands of real-time super-resolution rendering. In pathology, \gls{dicom}'s whole slide \gls{iod} provides a common format for integrating images into enterprise archives. However, this format does not natively support the pyramidal multi-resolution image structure. Instead, it stores individual tiles from a single \gls{wsi} pyramid level as separate frames within distinct \gls{dicom} multi-frame image files, with metadata duplicated and each tile's location explicitly encoded within its respective \gls{iod} \cite{Herrmann2018}. While, in our opinion, scanning vendors are correctly unifying behind the DICOM format as a long-term archive solution, there is a need for a simple vendor-agnostic transport format to be used in high-speed digital slide servers for digital sign-out \cite{Clunie2018}.

Modern real-time \gls{wsv} applications would greatly benefit from an intermediate ephemeral \gls{wsi} file format that can be rapidly encoded and accessed from a \gls{ims} and slide server instance or locally on the viewer workstation. Many contemporary performance-oriented \gls{wsv} software packages use an intermediate format. Viewers that use the OpenSeadragon library \cite{openseadragon}, such as Sectra IDS7 or Roche Navify, currently convert source \gls{wsi} files (such as \gls{dicom}) into \gls{dzi} format for daily use. This format is suboptimal because it consists of a markup file, such as \gls{xml}, along with a complex hierarchy of tile directories and subdirectories that are structured in a tree format based on resolution layers. Each directory contains individual image tiles, with each tile stored as a separate image file \cite{Schuffler2021}. Such solutions are cumbersome for enterprise quality implementations within modern healthcare environments and introduce the undesirable requirement of incorporating the operating system's file system as a critical internal component of the slide rendering software implementation. In an optimal design, the programmer should control the ``last mile'' of an \gls{ims} and thus retain programmatic control of the highest performance elements rather than delegating retrieval to the operating system in the form of repeated calls to the kernel's file system for basic data element lookup. Furthermore, large-scale hospital file management systems are agnostic to the functioning of the viewer system and, independent of \gls{ims} functioning, may fragment and cache directories across different storage media. 

\begin{figure}
    \centering
    \includegraphics[width=.8\linewidth]{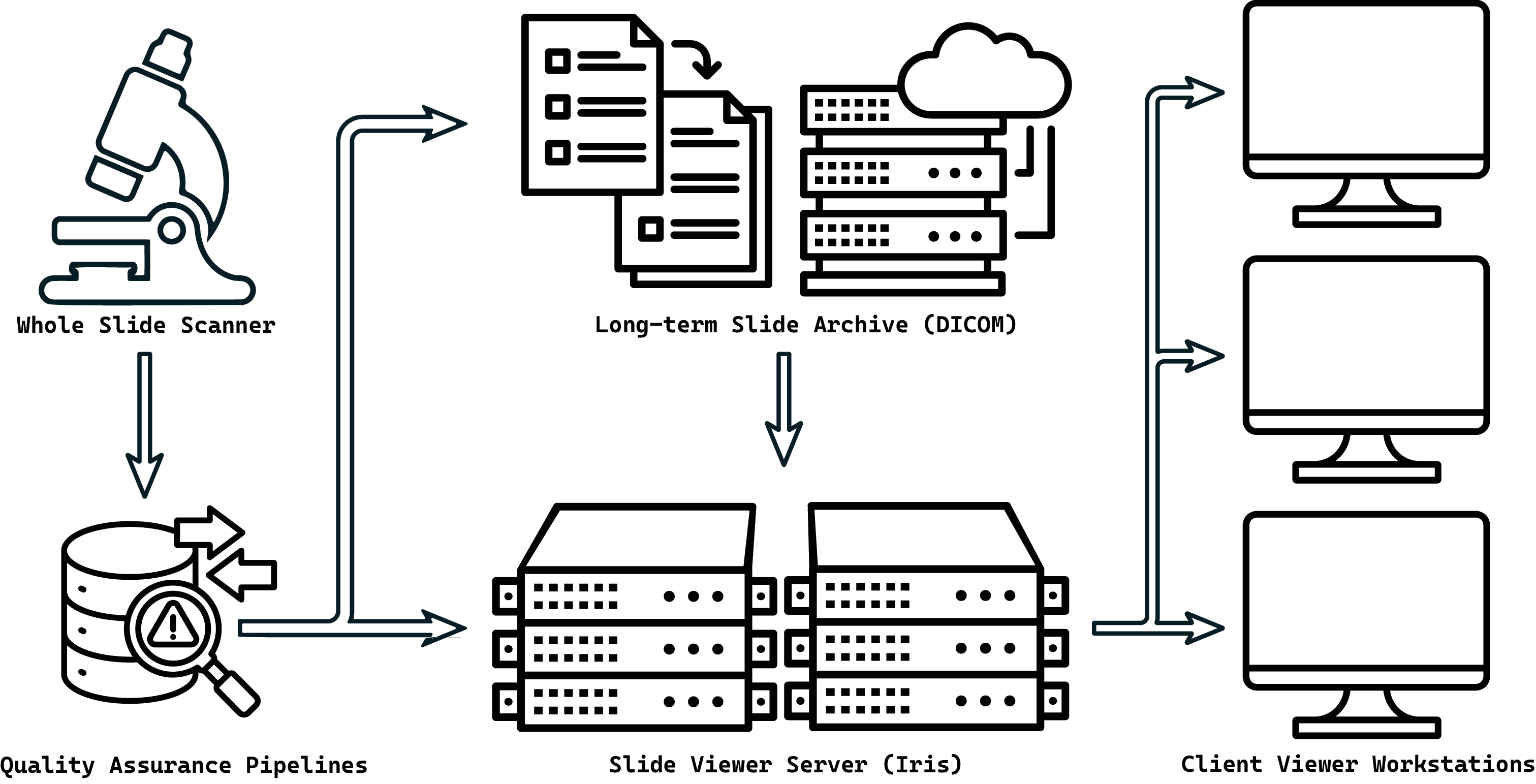}
    \caption{An example implementation of the Iris file extension file structure in a digital pathology workflow. \gls{wsi} data, following digital quality assurance pipelines, are stored as single high-resolution layers in \gls{dicom} within the long-term slide storage archives and are simultaneously converted to Iris files for performance viewer application access. If Iris files are purged from the viewer server, they may be unarchived and regenerated from \gls{dicom} storage, if needed.}
    \label{fig:workflow}
\end{figure}

In this article, we introduce the \Gls{ife} version 1.0 specification document (see supplement), which describes the file structure requirements and main goals for the Iris file extension transport format. We also provide the C++ source code and binary releases for all contemporary operating systems and architectures with language binding support for Python and JavaScript. Iris files are simple, binary, performance-oriented digital slides that complement \gls{dicom} in digital pathology workflows (Figure \ref{fig:workflow}). The Iris format serves as a temporary format by the \gls{ims} to accelerate image delivery and viewing on local workstations, generated during or after scanning, or when archived slides are retrieved. Iris files are not intended to replace \gls{dicom} as an interoperative long-term storage; rather, \gls{ife} provides a complementary structure to address the unique needs of \gls{wsi} rendering. Iris files may be purged when immediate access to a slide is no longer required, but they can be rebuilt with extreme speed using an unlimited number of parallel cores on demand. This allows \gls{dicom} files to be stored within a \gls{pacs} in a single flat high-resolution layer, saving space and avoiding current \gls{dicom} metadata duplication. The format is simple and explicit, optimized for data retrieval, and thus requires a reduced instruction set. This helps to enable client-side rendering at speeds comparable to those of glass slides.

\section{Design}
\begin{figure}
    \centering
    \includegraphics[width=0.75\linewidth]{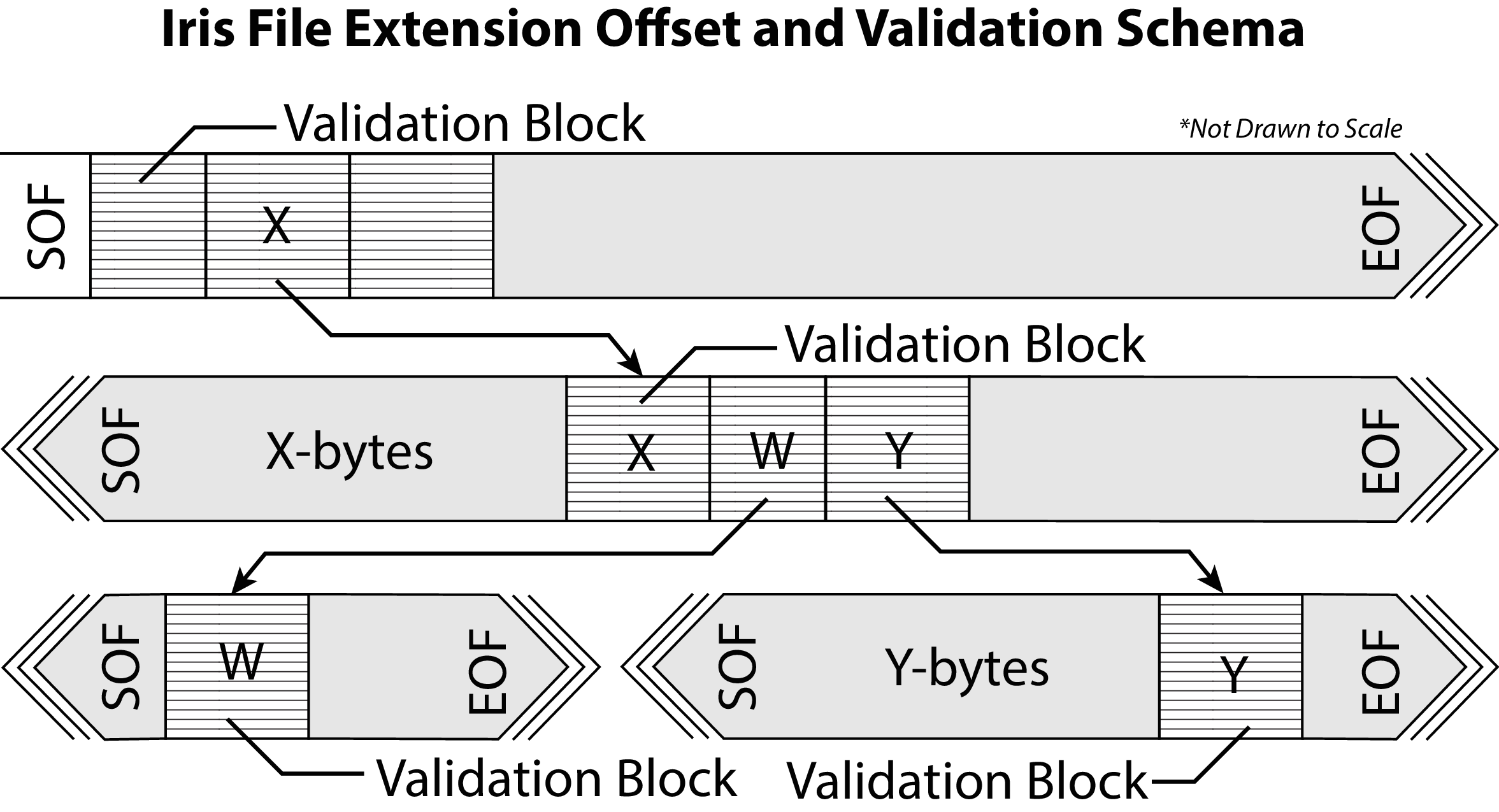}
    \caption{File offset linkage structure and validation scheme. Iris data-blocks are dynamically situated at encoder-defined 64-bit unsigned byte offset locations within the slide file. The 64-bit byte offset location is encoded at the beginning of any data-block such that a deserialize / read function $read(x)$ at the offset byte location $x$ evaluates as $read(x) = x$. In the above illustration, X is an unsigned 64-bit integer value. These offset pointers create the offset chain such that the valid data-block found at X points to the valid data-blocks located at W- and Y-bytes into the file.}
    \label{fig:offsetpointers}
\end{figure}

The IFE file structure comprises a series of tightly packed, structured \textbf{data-blocks}, akin to C-style structures, at dynamically defined byte offsets within the format byte stream. These data-blocks contain offset \textbf{validation tags} (Figure \ref{fig:offsetpointers}) that ensure file integrity and allow for deep but computationally trivial file validation routines. The specification achieves this by encoding the first 64 bits of any data-block with the data-block byte location relative to the start of the file. This validation tag ensures the integrity of the data-block offset-chain and is critical to file security; it ensures a completely safe and valid file before attempting \gls{wsi} tile reads. A system can recover the file structure if packet-loss or byte-corruption occurs within the data-blocks file structure using encoded data-block type-specific \textbf{recovery tags} that directly follow the validation tags. If file corruption occurs, a recovery system can scan for validation tags that encode their byte location, indicating the presence of a data-block, and then query the recovery tag to recover the data-block type.

Encoders can pack data-blocks in any location and any order, like \gls{tiff}. The \gls{ife} takes this a step further by engineering data-blocks for slide tile image lookup to support massively parallel writes efficiently. The developed structure allows for the simultaneous writing of tile data, even from different pyramid image layers, on an unlimited number of independently operating computer cores. We elaborate on this when describing the tile-table data-block below. The provided Iris Codec Encoder implementation source code provides an example of using an atomic byte offset counter to achieve this massively parallel encoding.

The \gls{ife} is strict in its requirements. The specification stipulates that all multi-byte values must be written in little-endian byte-ordering and all floating-point values in compliance with the \gls{ieee} and \gls{iec} established (IEEE-754/IEC-559) technical standard for floating-point arithmetic. In addition to promoting efficiency with most modern processors, this allows
 for backwards compatibility with 32-bit architectures and thus enables JavaScript to read \gls{ife} files natively (for files under the 32-bit limit) regardless of encoder architecture, unlike \gls{tiff}. The official implementations developed by the Iris Developers include open-source convenience functions that convert to and from little-endian byte ordering for big-endian systems, as well as convenience functions that allow conversion to and from serialized IEEE-754 floating-point values of any precision to aid in compliance with these requirements.

\begin{figure}[p]
    \centering
    \includegraphics[width=\linewidth]{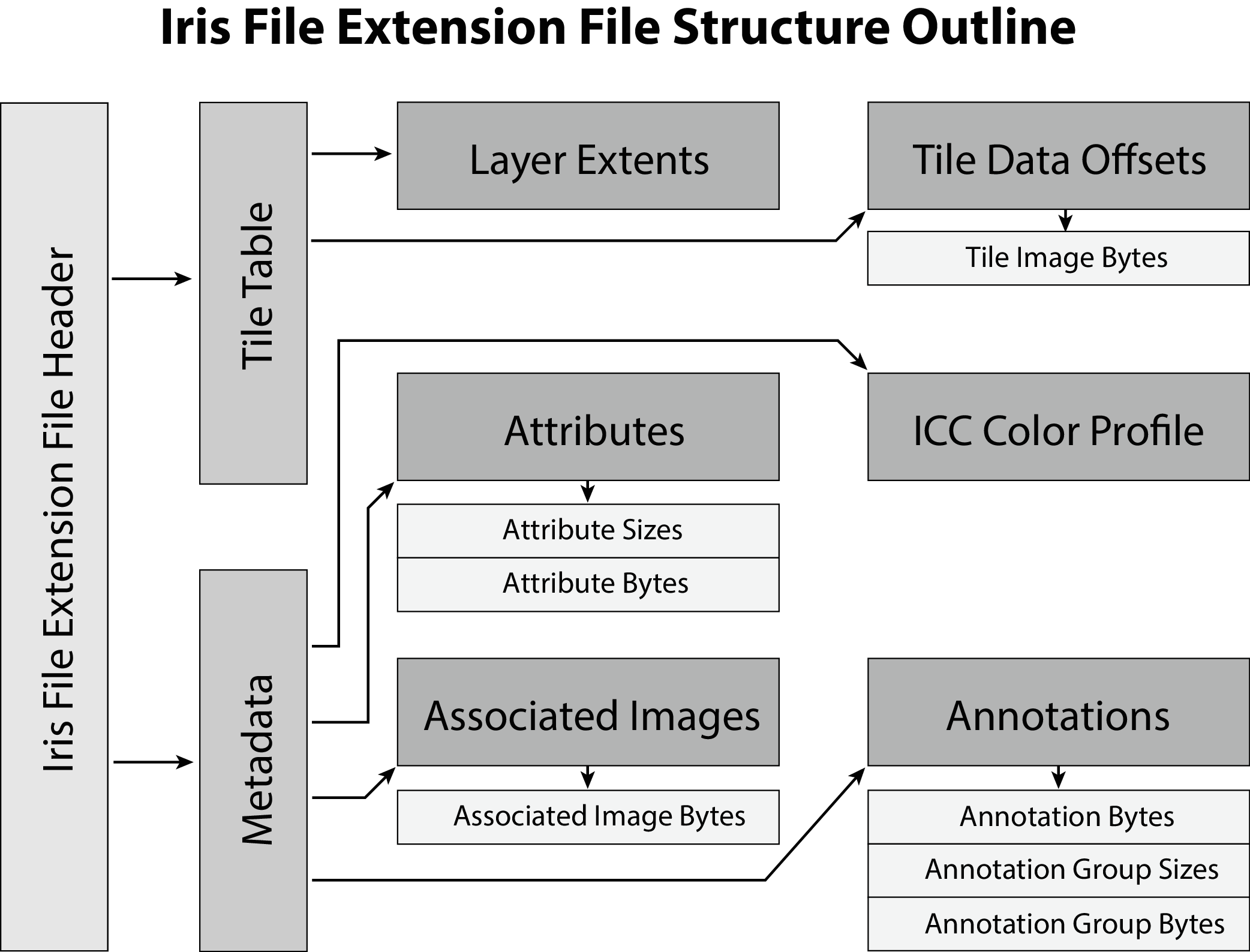}
    \caption{The Iris file structure follows a series of linked data-blocks that extend four layers deep. More data-blocks may be added in future revisions and more layers may be added to the file depth if needed. The file-header points to tile-table and clinical-metadata blocks. The tile-table points to the layer-extents and tile-offsets array, which describes where to read serialized tile data (Figure \ref{fig:tiletable}). In the same layer, the metadata points to attributes, associated-images, \gls{icc} color profile, and annotations array. Many of these tertiary data-blocks are array types and point to data-blocks wrapping serialized byte arrays (white boxes).}
    \label{fig:fileoutline}
\end{figure}

The complete data-block file structure follows a series of linked byte offset locations between data-blocks, referred to as the \textbf{offset-chain} (visually outlined in Figure \ref{fig:fileoutline}). The file structure data-blocks include \textbf{header-blocks} (\gls{ife} Specification Section 2.3) and \textbf{array-blocks} (\gls{ife} Specification Section 2.4). There is little difference between array blocks and header-blocks, except that array blocks contain a dynamically sized array of entries as their last parameter. Importantly, the size of an array entry may grow in later versions of the \gls{ife} specification, but prior arrays remain completely valid in the specification. This allows for bidirectional (including future edition) compatibility of the file structure. Arrays store the array entry byte size at the time of encoding, ensuring proper read alignment regardless of system or file specification version (past or future). The complete file structure will be described below.

The \textbf{file header} appears first in the file, providing initial validation information and the byte offsets to the \textbf{tile-table} and \textbf{clinical-metadata} data-blocks. The tile-table contains all necessary information for a \gls{wsv} application to render the slide view, including all necessary information to decode compressed image byte streams and format the \gls{wsi} pyramid structure. The clinical-metadata data-block contains all ancillary information, such as file attributes containing scanning, clinical, and histologic information. This data-block also grants access to key information, including the \gls{icc} profile, associated images such as slide labels and thumbnails, and digitally encoded slide annotations and annotation groupings, all embedded within the \gls{ife} file.

\begin{figure}[t]
    \centering
    \includegraphics[width=\linewidth]{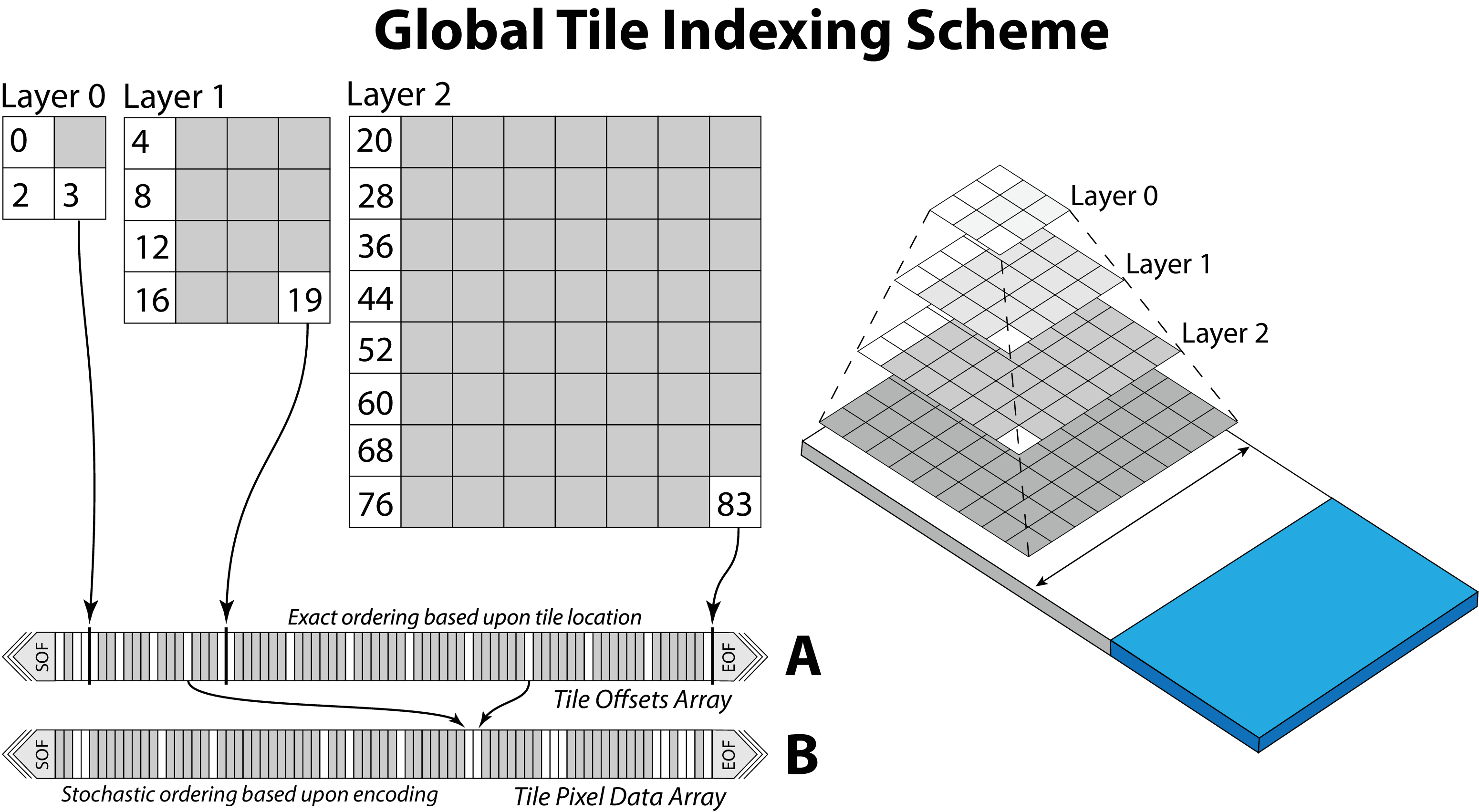}
    \caption{Global Tile Indexing Scheme. Tiles are packed by layer with a unique index value assigned to each tile, regardless of layer. The indexing of layers is arranged from lowest scale, on the \gls{sof} side, to highest scale towards the \gls{eof}. Tiles within layers are arranged with upper left tiles towards the \gls{sof} and bottom right towards the \gls{eof}. The \textbf{tile-offset} array (A) makes use of this global indexing scheme while the packing of \textbf{tile pixel data} (B) may be arranged in any order the encoder chooses, which can be entirely stochastic.}
    \label{fig:globalindexing}
\end{figure}

\begin{figure}
    \centering
    \includegraphics[width=\linewidth]{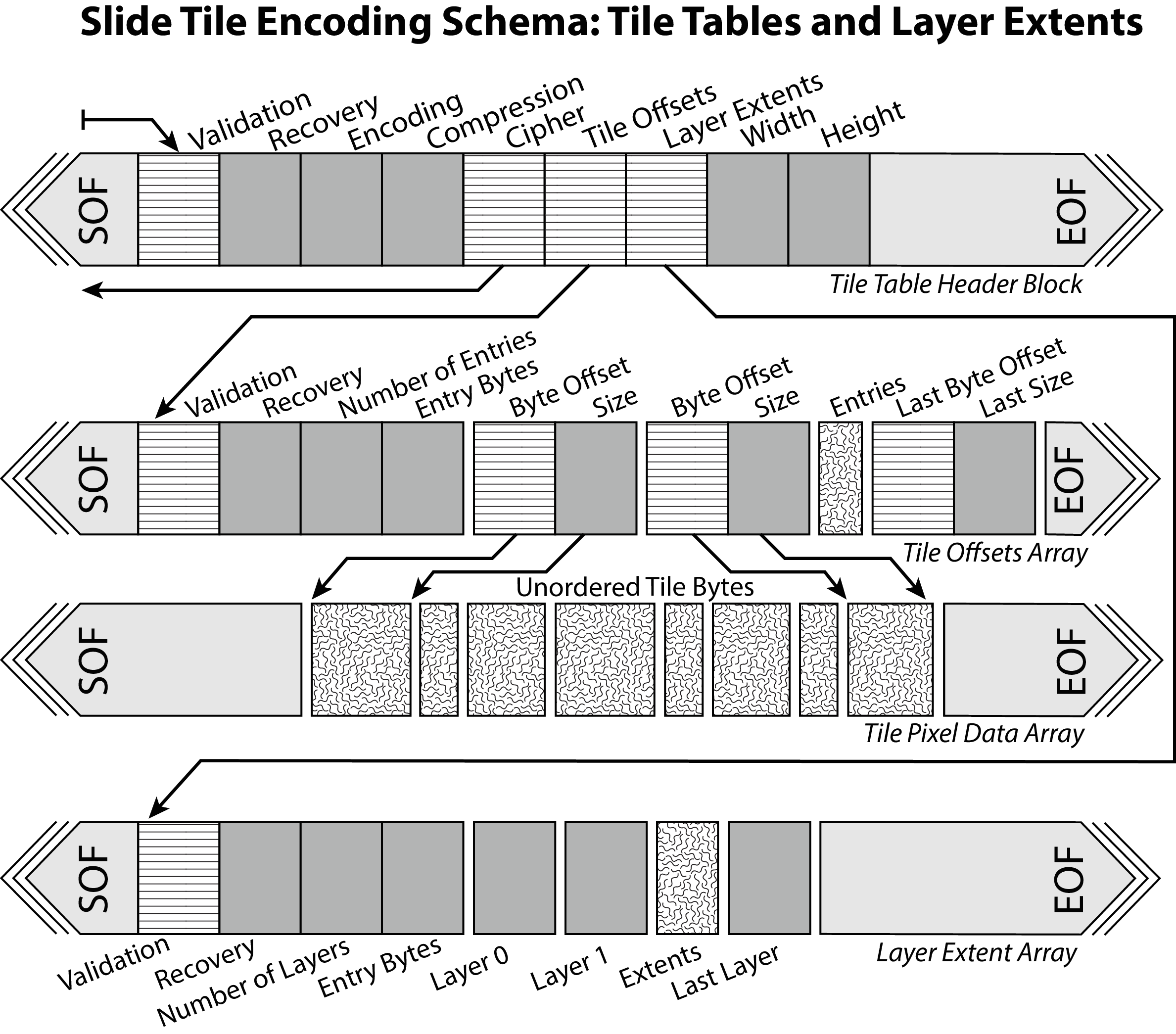}
    \caption{Slide tile encoding scheme detailing the offset trees of the tile-tables and tile arrays, and the slide / layer extents. Offset parameters are shown as horizontal hash marks while value parameters are shown as gray boxes. Variable sized or a variable number of entries are illustrated with a texture. The \textbf{tile-table} points to a \textbf{tile-offset} array and a \textbf{layer-extents} array. The \textbf{tile-offsets} array encodes the offset and size in bytes of the each variably sized slide \textbf{tile pixel data} array. The layer-extent array lists parameters such as the number of tiles in each dimension and relative scale of each layer encoded in the file.}
    \label{fig:tiletable}
\end{figure}

The \textbf{tile-table} header data-block contains all the essential information needed to render the slide. This includes the encoding format for selecting the appropriate image compression codec, the original pixel format to determine bit depth and alpha components, and offset pointers to the layer-extents and tile-offsets array data-blocks. These array data-blocks define the pyramid structure and provide a multidimensional index for efficient random access to slide tiles. The currently supported image codecs include the \gls{jpeg} format and the \gls{avif}. We outline our rationale for these image codecs in the discussion section. The layer-extents array provides the layer dimensions in tiles for each pyramid layer and that layer's relative scale. Using the layer-extents array, a decoder can derive a \textbf{global tile indexing scheme} (Figure \ref{fig:globalindexing}) that reduces the multi-dimensional tile index of layer-, x-, and y-coordinates into a single dimensional array that decoders can use to look up the byte location of any slide tile at any resolution layer. The tile-offsets array encodes the byte offset locations and compressed sizes of all slide tile data, and is ordered according to the global tile indexing scheme. Sparse tiles are permitted and are encoded with a sparse-tile value. While an encoder may write compressed tile data to the slide file byte stream in any order, the tile-offsets array must be ordered according to the global tile indexing scheme. The full serialization structure for the tile-table and derivative data-blocks for version 1.0 of the specification is outlined visually in Figure \ref{fig:tiletable}.

\begin{figure}[p]
    \centering
    \includegraphics[width=\linewidth]{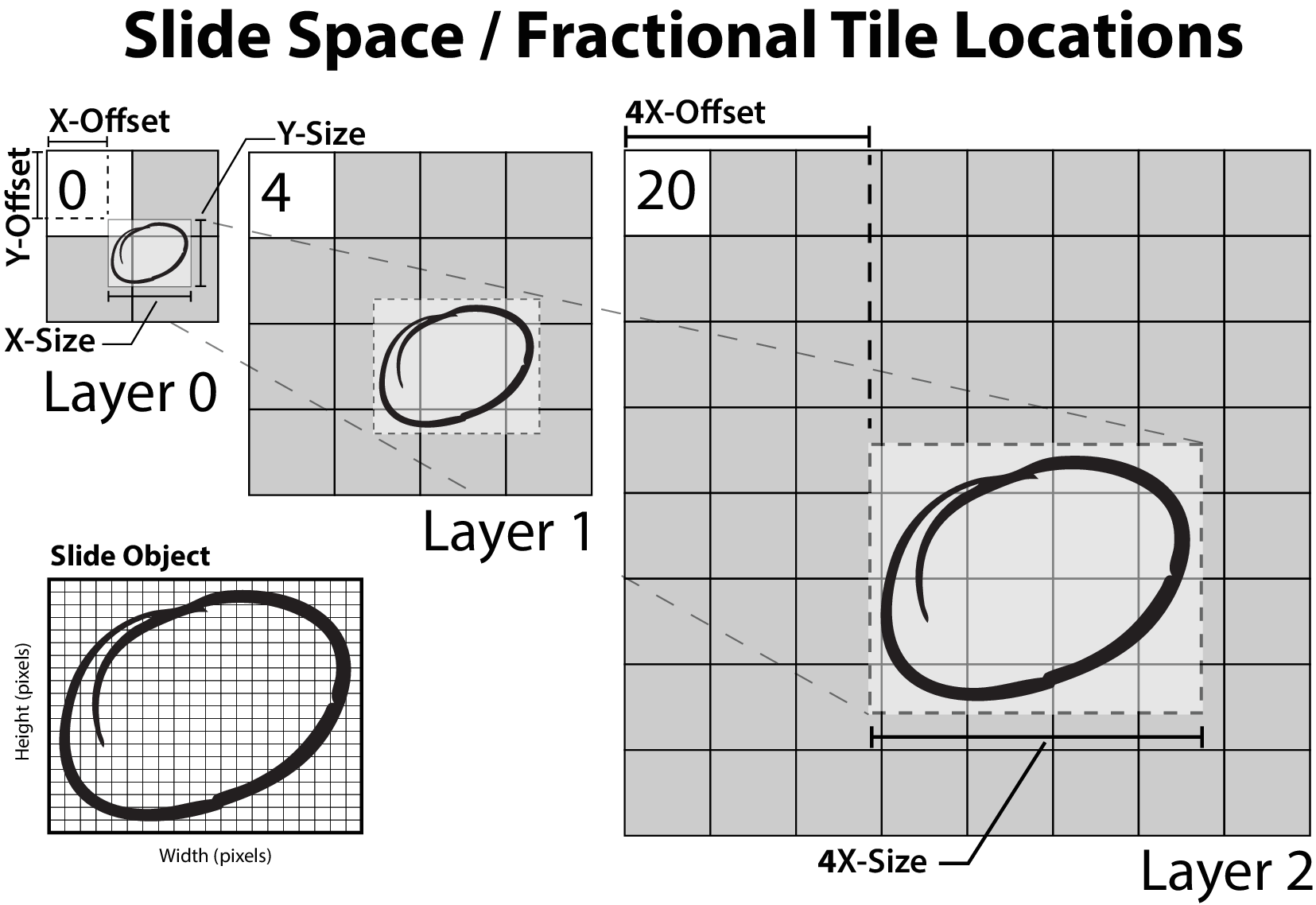}
    \caption{Slide objects are graphical figures, in the form of a raster graphic (that may optionally be derived from vector graphics) with the height and width dimensions in pixels (bottom-left). The rendered dimensions are separate from the raster dimensions, as they must be mapped to the slide within slide space, a coordinate notation that maps well to digital slides' multi-resolution layers. The slide space is defined in floating-point tile numbers at the lowest-resolution layer (0). This range for each axis is defined as [0.0, number of tiles], and in the above example, it would be [0.0, 2.0] in both x- and y-dimensions. This space may be transformed based on the viewer's zoom / resolution by simply multiplying by the current zoom amount. The location of an annotation is represented in terms of the offset and size. In the above implementation, the X-offset is approximately 0.73 tiles with an X-size of approximately 1.0 tiles at the initial layer (0). When drawn using a whole-slide image viewer, such a structure can be appropriately scaled with ease as the viewer changes scale (right).}
    \label{fig:annotations}
\end{figure}

The \textbf{clinical-metadata} header block is a required data-block element that contains a series of optional arrays allowing additional information about the slide file, patient, or slide annotations. The metadata may reference an optional attributes data-block array, which stores key-value pairs. These pairs are encoded either according to the \gls{dicom} PS3.3 standard for full \gls{dicom} compatibility \cite{ddicomps3} or as \gls{ascii} keys with \gls{utf8} values. We have included the option to map entirely to the \gls{dicom} specification for full interoperability with the \gls{dicom} standard. We are also expanding upon the PS3.3 dictionary in an interoperability specification known as the \gls{i2s}, which we will elaborate on in the discussion section. The metadata header block may also point to an optional \gls{icc} color profile data-block to properly render the slide images in the intended color space and \textbf{associated images} data-block array (such as thumbnail images, slide label images, etc.) that are labeled according to their intended use. The final data-block in the clinical-metadata offset-chain is the \textbf{annotations} data-block array (Figure \ref{fig:annotations}), which may contain digitized on-slide annotations.

The annotations data-block array offers a simple, interoperable mechanism for encoding on-slide annotations by pathologists. We presently support text annotations and support vector and rasterized graphical formats. Annotations are applied to slides at layer-independent floating locations termed \textbf{fractional tile locations} within a layer-agnostic coordinate system (termed \textbf{slide space}) that are interoperable with the modern graphical \gls{api} coordinate system and the previously described global indexing scheme (Figure \ref{fig:globalindexing}). The annotation size in slide space is entirely independent from the raster size of the graphical annotation object, which must be separately encoded as pixel width and height. The raster size is either the intended rasterized size (for vectors such as text or scalable vector graphics) or the encoded dimensions of compressed image byte-streams. All annotations on a digital slide must be assigned a unique \textbf{annotation identifier} for viewer systems to reference the specific annotation. Further, this annotation identifier can be used to create linked annotations in a parent-child relationship and generate named annotation groups, which are collections of annotations with a given theme. Annotation groups are particularly important in developing digital pathology classifier systems.

\section{Implementation}
\subsection{Releases}
The \gls{ife} is released as a digital slide file extension specification. We have made this source code available for community and commercial applications, and we use the \gls{ife} internally in our development of high-performance digital slide rendering systems and compression algorithms. We have also released pre-compiled binaries for the native implementation in C++ with language bindings for Python, via the Anaconda Python package manager in the Conda-Forge channel, and JavaScript using Emscripten-compiled \gls{wasm} binaries. We have extensively evaluated the system on a variety of architectures and operating systems with effective support for Linux (x86-64 / AMD64, aarch64 / arm64, and arm64v8 architectures), macOS (x86-64 and arm64), iOS (arm7 and arm64), and Windows (x86-64).

The \gls{ife} may be implemented in two ways: (1) We provide a low-level Iris-File-Extension direct serialization/deserialization and specification validation library, and (2) we provide a higher-level abstraction \gls{api} that performs most low-level image compression and serialization routines behind the scenes as part of the larger Iris Codec compression module. The Iris Codec Community Module depends on the Iris-File-Extension library. Projects using the \gls{ife} can include one or both of these libraries, allowing customization of the file structure to optimize existing workflows as needed.

Option (1) is the Iris-File-Extension \gls{api}, a serialization and validation library. It does not incorporate compression codecs nor file input/output calls to the operating system kernel. This library is more hands-off and is intended for slide scanning vendors or programmers who wish to develop custom encoders or decoders that adhere to the file specification. It may be used to perform deep validations of existing \gls{ife} slide files, de-serialize data in an optionally zero-copy manner, and write specification-validated data-blocks to a specific offset location within a slide file. Most programmers who wish to use the \gls{ife} should not use this library; instead, they should explore option (2). A complete and up-to-date explanation of installation and use with and without the \gls{ife} file abstraction assistance is available on the \href{https://github.com/IrisDigitalPathology/Iris-File-Extension/blob/main/README.md}{online repository documentation}.

Option (2) is the Iris Codec Community Module, which is a submodule of the larger Iris Digital Pathology project that manages compression routines and access to \gls{ife} encoded slide files. Following validation, this module ingests the \gls{ife} file structure and generates a file abstraction for rapid slide data retrieval. Unlike the Iris-File-Extension library, the codec community module is a full-feature drop-in implementation that outputs fully decompressed slide tile data. It also integrates with the Iris Core rendering module for GPU-based codec acceleration when supported. This is made possible by the image codec's compatibility with the officially supported tile encoding types in the \gls{ife} specification. Presently, these depend upon Turbo-Jpeg and \gls{avif}. However, support for the Iris Codec compression algorithm (our ongoing research focus) for slide image data will be added in future releases. Additional image codecs may also be added based upon community requests (such as the H.265 for backwards compatibility with mainstream hardware at the time of this publication).

The codec community module exposes an object-oriented \gls{api} that closely matches the \gls{ife} specification. Slide objects are created from \gls{ife} file paths and are initialized as a transparent, nested structure that maps well to modern languages like Python. This structure contains all relevant slide metadata while keeping binary data on disk, loading it on-demand. Slide objects include \gls{wsi} tile image data, associated images, annotations, and \gls{icc} color profiles. For example, the clinical-metadata data-block holds a list of associated-image labels or annotation identifiers in its nested structure. A label or annotation identifier may then retrieve the specific associated image or annotation using the \gls{api}'s various 'read' functions. This provides a lightweight and easily navigable object-oriented file structure in-memory representation, while delivering quick on-demand access to larger data payloads.

The codec community module also provides encoder routines as part of the underlying \gls{api} and as a separate command line executable. Iris slide files comprise 256 x 256 pixel tiles. When the input file is similarly structured, the encoder copies an unaltered compressed byte stream to the destination \gls{ife} container to avoid loss in image quality. Alternatively, for vendor files of alternative formats supported by OpenSlide,  the encoder may use OpenSlide to read raw pixel values. Note, if the encoder is built with the optional OpenSlide dependency, its licensing agreement changes to GPLv2 in compliance with OpenSlide's license. The Iris codec community encoder routine is designed to utilize a system's processor cores fully and can encode most files in 1-2 minutes, even when using OpenSlide.

As with the Iris-File-Extension library, please refer to the \href{https://github.com/IrisDigitalPathology/Iris-Codec/blob/main/README.md}{online repository documentation} for a complete and up-to-date explanation detailing installation and full use of the Iris Codec Community module.

\subsection{Access Performance}

 \gls{ife} performance is derived from the specification's emphasis on simplicity, which results in minimal instructions required for data retrieval. Read access was profiled on a 2020 13-in M1 MacBook Pro (Apple, Palo Alto, CA) with 8 GB of RAM running macOS 15.4 using a Python script in batches of 10,000 tiles (Python 3.11; Anaconda environment) for local \gls{dicom} and \gls{ife} files, and using the Chrome browser JavaScript developer tools for DICOMweb over a gigabit connection. The imi-bigpicture WsiDicom \cite{WsiDicom} python package by Erik Gabrielsson, SectraAB, was selected because it was the highest performance \gls{dicom} de-serializing package we could identify. \gls{dicom} examples were sourced from the NCI \gls{idc} \cite{Fedorov2021} from three separate collections (CCDI-MCI, CPTAC, and CMB-BRCA). The exact study instance \glspl{uid} are provided in Table \ref{tab:uids}. Tiles were retrieved as decompressed images from the highest resolution layer.
 
\begin{figure}[t]
    \centering
    \includegraphics[width=1\linewidth]{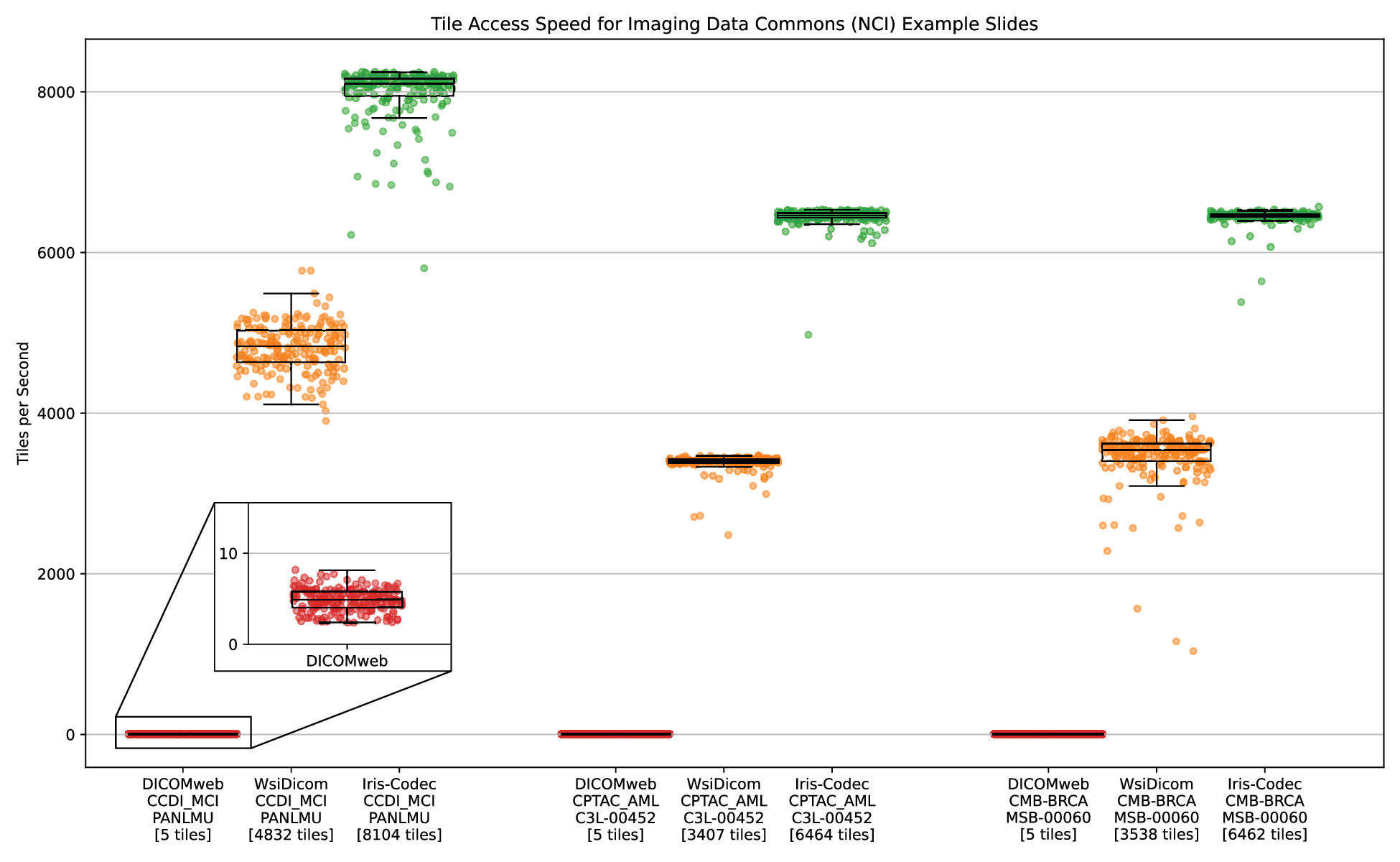}
    \caption{Tile access speeds for the National Cancer Institute Imaging Data Commons \gls{wsi} example slide files. Read rates for three datasets (CCDI-MCI, CPTAC, and CMB-BRCA; Table \ref{tab:uids}) are compared between the DICOMweb protocol from the Slim \cite{Gorman2023} image viewer and local slide files in DICOM and IFE formats using WsiDicom \cite{WsiDicom} and the Iris-Codec implementations. Median number of tiles retrieved per second (access speed) is noted below, each in square brackets.}
    \label{fig:performance}
\end{figure}

 Tile access rates were, on average, approximately 5 tiles/sec for Slim/DICOMweb, 3400-4800 tiles/sec for WsiDicom, and 6400-8100 tiles/sec for Iris-Codec (Figure \ref{fig:performance}). These correspond to \gls{tpt} values of approximately 200 ms for DICOMweb, 0.25 ms for WsiDicom, and 0.14 ms for Iris-Codec. These \gls{tpt} rates for Iris-Codec are consistent with our previously reported buffering rates of 0.10 - 0.16 ms/tile during Iris rendering \cite{Landvater2025}.

\section{Future Work}

The next edition of the \gls{ife} specification will incorporate the Iris File Extension Networking Protocols for server-hosted slide tile data. These protocols are already used internally within our Iris Networking module; however, they must be thoroughly tested and refined before being formally codified in the specification document. Performance metrics for slide encoding rates and file recovery/resilience are outside the scope of this initial specification publication and will be reported in later works.

A major current focus is developing the Iris Codec compression algorithm to improve slide tile compression yields. While this is not within the scope of the \gls{ife}, the extension was originally developed as a container to house Iris Codec-encoded slide data and has reserved constants within the specification for its implementation. We also plan to add GPU-based AV1 encoder support in subsequent releases of the Iris Codec Community module by incorporating the newly released Vulkan KHR Video Decode AV1 extension, as well as Apple's Video Toolkit framework. We may add additional compression codec support, such as JPEG2000/XL or H.265 (HEVC), if there is widespread community interest in their inclusion.

\section{Discussion}

The \gls{ife} is designed to be simple, versatile, and fast. The \gls{ife} is a high-performance binary container format amenable to massively multithreaded file encoding writes and immediate multithreaded random access. We designed the \gls{ife} structure to expand upon the most convenient features of \gls{tiff}, a familiar image format employed in digital pathology \cite{Clunie2019}, but redesigned for use in viewer-oriented digital pathology workflows. We have incorporated advancements in language-neutral performance serialization technology to influence the dynamic structure of the \gls{ife}. Similarities to \gls{tiff} and dynamic information ordering enable the \gls{ife} structure to seamlessly integrate with established production workflows. The \gls{ife} is designed to be simply ``dropped in'' to existing pipelines and workflows, allowing them to immediately begin taking advantage of the \gls{ife}'s expanded capabilities. Additionally, we added easily configurable optional data-blocks and data ordering that allow for unobtrusive file updates when encoders preferentially place dynamic information near the end of the file.

The specification is highly explicit in byte structuring, even concerning dynamic data elements at encoding time. This allows for complete bidirectional file and encoder version compatibility. An earlier version of a decoder can safely access future \gls{ife} encoded files, even if an old decoder is unaware of the added features a more modern slide file may contain. Further, this design paradigm has led to a very secure file format that validates structures and allows data recovery if file structure metadata is ever corrupted. 

We describe but do not fully define a new clinical-metadata attribute interoperability specification, the \gls{i2s}. This specification is not meant to supplant the \gls{dicom} PS3.3 attributes \cite{ddicomps3} that follow the \gls{dicom} tag-value standard; rather, we hope this forthcoming unicode dictionary aids in the role of Iris files as short-term and interoperable file containers for \gls{wsi} outside of the definitive archives. Additional information for daily-use, high-performance viewer pipelines, vendor specific viewer pipeline information, or other data not yet incorporated into the \gls{dicom} standard may be needed within workflows. The \gls{i2s}  will provide a standardized set of key-values such that this information can be transcribed into a slide file while ensuring inter-institutional standardization. This will be further described in a future publication. 

We designed the \gls{ife} structure with forward compatibility and digital image media advancements at its core. Ongoing development incorporates the future of digital imaging technology within pathology informatics, and we plan to continue developing this standard by aiming for where the field will be, rather than focusing on its current state. For example, the \gls{ife} not only supports wrapping \gls{jpeg}, the de-facto \gls{wsi} file standard, but also \gls{avif} due to emerging GPU-based encoder hardware for the AOM video 1 compressed image stream. These hardware encoders are being offered by hardware manufacturers like \href{https://nr.apple.com/Di5I4t7da8}{Apple} \cite{appleAV1} and \href{https://docs.nvidia.com/video-technologies/video-codec-sdk/13.0/read-me/index.html}{Nvidia} \cite{nvidiaAV1} and have been formally incorporated in official graphical \glspl{api}, such as \href{https://khr.io/12p}{Vulkan} \cite{vulkanAV1}. This enables high-performance rendering systems, such as Iris Core, to transfer compressed tile byte streams to the GPU and render directly from the hardware decoder outputs. The \gls{ife} aims to leverage advancements in highly innovative spheres, such as the video-streaming industry, to provide \gls{ife}-derived implementations with access to cutting-edge encoders for the fastest and most energy-efficient slide viewer applications.

In previous work \cite{Landvater2025}, we have shown that the Iris Core rendering module can buffer and render to screen an entirely new \gls{fov} in less time (TeFOV) than a single standard display frame refresh using the Iris format (TeFOV 10-35 ms; TPT 0.10-0.16 ms/tile) -- achieving a user experience comparable to moving a glass slide under a microscope \cite{Landvater2025}. This high-speed performance, even for large whole-slide images, enables pathologists to navigate and zoom with zero perceptible lag, thereby closing the usability gap between digital and optical workflows.

We have compared these with the Slim viewer system \cite{Gorman2023} that renders DICOM files natively using the DICOMweb RESTful \gls{api}; however, this is not a direct comparison of system-to-system as the latter introduces network transmission as a variable. High-performance network implementations have been made by Schuffler et. al. using their FlexTileSource, and these showed full FOV rendering in 240 ms \cite{Schuffler2022}. Performance metrics for Slim, both local rendering performance and network rendering metrics have not been published. In future work, we will report network performance of the IFE when the networking module is available.

Crucially, Iris is an ephemeral format. The authoritative image record remains a single high-resolution \gls{dicom} version stored in the archival system, which retains all metadata and compliance with long-term management standards. In our proposed model (Figure \ref{fig:workflow}), \gls{dicom} continues to serve as the indelible reference and interoperability layer, while Iris serves as a nimble intermediate layer geared toward simplicity and \gls{wsi} system performance. This separation of roles leverages the strengths of both formats: the universality and rich metadata support of \gls{dicom} and the raw performance of Iris.

We hope that successful demonstration of the \gls{ife} as a performance-oriented transport format within a \gls{dicom}-based \gls{ims} could prove the utility of the simplified Iris byte structure, specifically the tile-table data-blocks (Figure \ref{fig:tiletable}), as viewer rendering oriented byte-stream. We hope that the \gls{dicom} specification committees continue to watch \gls{ife} use for possible consideration of \gls{ife} elements as an \gls{iod} within future versions of the \gls{dicom} specification. Such incorporation of the \gls{ife} byte stream elements into the \gls{dicom} standard at a later date could ease future transition from a performance-oriented intermediate format to direct rendering from archive files for enterprise implementations that already take advantage of the Iris-Codec's limited instruction set and streamlined API. 

\section{Conclusion}

The \gls{ife} is a high-performance binary container format for whole slide images, designed as an intermediate, ephemeral format to enable high-speed \gls{wsi} file server operations and efficient local slide rendering on client machines. The format enables rapid random-access reads and parallel writes in a manner that delegates the dynamic byte structure to encoder implementations. This allows it to be easily incorporated into existing workflows. It has an explicitly defined byte format that allows bidirectional file-encoder compatibility through computationally trivial deep validation routines. The file structure enables recovery in the event of critical file structure corruption due to redundant information, including validation and recovery tags. We provide this file extension to the community in the form of a specification document (licensed under the Creative Commons Attribution-No Derivative 4.0) and as drop-in C++ source code (under the MIT software license) or binary builds with Python and JavaScript language bindings.

\section{Acknowledgements}
We have no additional funding sources to disclose. We would like to acknowledge Drs. Corey Post, MD and Vincent Laufer, MD PhD for thorough manuscript review and recommendations. We would like to acknowledge Kenton Varda, the author of Protobuffers and Cap'n Proto serialization libraries. His responses were valuable in the design of the Iris file format. Further, methodology in Iris' serialization routines are loosely based upon Kenton Varda's Protobuffer and Wouter van Oortmerssen's Flatbuffer serialization code base. 

\section{Author Contributions}

RL designed and developed the Iris File Extension file structure, drafted the file structure specification document, wrote the C++ implementation and Python binding code, wrote the Design and Implementation sections, and contributed to the other sections. MO wrote the JavaScript binding code and contributed to all manuscript sections. MY contributed clinical insights for elements of the file structure design. UB provided guidance throughout the development of the file structure and contributed to the Introduction and Discussion sections. All authors reviewed and approved the final manuscript. 

\section{Data Availability Statement}
Implementations are available from the \href{https://github.com/IrisDigitalPathology}{Iris Digital Pathology} Github organization page (\url{https://github.com/IrisDigitalPathology}). Source code is provided for the \href{https://github.com/IrisDigitalPathology/Iris-File-Extension}{Iris-File-Extension} \gls{api} library. The source code is also available for the \href{https://github.com/IrisDigitalPathology/Iris-Codec}{Iris Codec Community Module}, which contains (de)serialization and (de)compression / encoding routines to access \gls{ife} encoded files, using a simplified \gls{api}. The Iris Codec Community module is also provided as \href{https://anaconda.org/conda-forge/iris-codec}{Python language bindings}, hosted on the Conda-Forge channel of the Anaconda package manager at \url{https://anaconda.org/conda-forge/iris-codec} and PiP package repository at \url{https://pypi.org/project/Iris-Codec}. Example \gls{wsi} files encoded using the \gls{ife} are hosted on the \href{https://github.com/IrisDigitalPathology/Iris-Example-Files}{Iris-Example-Files} repository. The python script used to execute performance testing is included in the supplemental materials and the study \glspl{uid} used are provided in Table \ref{tab:uids}.
\begin{table}[h]
    \centering
    \begin{tabular}{l l l}
         Collection & Case & \gls{uid}  \\\hline
         CCDI-MCI & PANLMU &2.25.60737598245052570577932078803929433012 \\
         CPTAC-AML & C3L-00452 & 2.25.228065630334552244523047308715937146144 \\
         CMB-BRCA & MSB-00060 & 2.25.166829711238260757039099289902569655832 \\
    \end{tabular}
    \caption{NCI \gls{idc} Study Instance \glspl{uid} used in read access testing}
    \label{tab:uids}
\end{table}


\appendix
\section{Glossary of Terms}
\printglossaries

 \bibliographystyle{elsarticle-num}
 \bibliography{bibliography.bib}

\begin{thebibliography}{10}
\expandafter\ifx\csname url\endcsname\relax
  \def\url#1{\texttt{#1}}\fi
\expandafter\ifx\csname urlprefix\endcsname\relax\def\urlprefix{URL }\fi
\expandafter\ifx\csname href\endcsname\relax
  \def\href#1#2{#2} \def\path#1{#1}\fi

\bibitem{Landvater2025}
R.~E. Landvater, U.~Balis, \href{https://linkinghub.elsevier.com/retrieve/pii/S2153353924000531}{Iris: A next generation digital pathology rendering engine}, Journal of Pathology Informatics 16 (2025) 100414.
\newblock \href {https://doi.org/10.1016/j.jpi.2024.100414} {\path{doi:10.1016/j.jpi.2024.100414}}.
\newline\urlprefix\url{https://linkinghub.elsevier.com/retrieve/pii/S2153353924000531}

\bibitem{Schuffler2022}
P.~J. Sch{\"{u}}ffler, E.~Stamelos, I.~Ahmed, D.~K. Yarlagadda, O.~Ardon, M.~G. Hanna, V.~E. Reuter, D.~S. Klimstra, M.~Hameed, {Efficient Visualization of Whole Slide Images in Web-based Viewers for Digital Pathology}, Archives of Pathology and Laboratory Medicine 146~(10) (2022) 1273--1280.
\newblock \href {https://doi.org/10.5858/arpa.2021-0197-OA} {\path{doi:10.5858/arpa.2021-0197-OA}}.

\bibitem{Goode2013}
A.~Goode, B.~Gilbert, J.~Harkes, D.~Jukic, M.~Satyanarayanan, \href{https://linkinghub.elsevier.com/retrieve/pii/S2153353922006484}{Openslide: A vendor-neutral software foundation for digital pathology}, Journal of Pathology Informatics 4 (2013) 27.
\newblock \href {https://doi.org/10.4103/2153-3539.119005} {\path{doi:10.4103/2153-3539.119005}}.
\newline\urlprefix\url{https://linkinghub.elsevier.com/retrieve/pii/S2153353922006484}

\bibitem{Clunie2021}
D.~A. Clunie, Dicom format and protocol standardization—a core requirement for digital pathology success, Toxicologic Pathology 49 (2021) 738--749.
\newblock \href {https://doi.org/10.1177/0192623320965893} {\path{doi:10.1177/0192623320965893}}.

\bibitem{Herrmann2018}
M.~D. Herrmann, D.~A. Clunie, A.~Fedorov, S.~W. Doyle, S.~Pieper, V.~Klepeis, L.~P. Le, G.~L. Mutter, D.~S. Milstone, T.~J. Schultz, R.~Kikinis, G.~K. Kotecha, D.~H. Hwang, K.~P. Andriole, A.~J. lafrate, J.~A. Brink, G.~W. Boland, K.~J. Dreyer, M.~Michalski, J.~A. Golden, D.~N. Louis, J.~K. Lennerz, Implementing the dicom standard for digital pathology, Journal of Pathology Informatics 9 (2018) 37.
\newblock \href {https://doi.org/10.4103/jpi.jpi_42_18} {\path{doi:10.4103/jpi.jpi_42_18}}.

\bibitem{Clunie2018}
D.~Clunie, D.~Hosseinzadeh, M.~Wintell, D.~D. Mena, N.~Lajara, M.~García-Rojo, G.~Bueno, K.~Saligrama, A.~Stearrett, D.~Toomey, E.~Abels, F.~V. Apeldoorn, S.~Langevin, S.~Nichols, J.~Schmid, U.~Horchner, B.~Beckwith, A.~Parwani, L.~Pantanowitz, Digital imaging and communications in medicine whole slide imaging connectathon at digital pathology association pathology visions 2017, Journal of Pathology Informatics 9 (2018) 6.
\newblock \href {https://doi.org/10.4103/jpi.jpi_1_18} {\path{doi:10.4103/jpi.jpi_1_18}}.

\bibitem{openseadragon}
I.~Gilman, Openseadragon 5.0.1, \url{https://openseadragon.github.io}, accessed: April 4, 2025.

\bibitem{Schuffler2021}
P.~J. Schüffler, G.~G. Ozcan, H.~Al-Ahmadie, T.~J. Fuchs, Flextilesource: An openseadragon extension for efficient whole-slide image visualization, Journal of Pathology Informatics 12 (2021) 31.
\newblock \href {https://doi.org/10.4103/jpi.jpi_13_21} {\path{doi:10.4103/jpi.jpi_13_21}}.

\bibitem{ddicomps3}
{DICOM Standards Committee}, Dicom ps3.3 2025a - information object definitions, \url{https://dicom.nema.org/medical/dicom/current/output/html/part03.html}, accessed: April 4, 2025 (2025).

\bibitem{WsiDicom}
E.~O. Gabrielsson, Wsidicom 0.27.1, \url{https://github.com/imi-bigpicture/wsidicom}, accessed: May 26, 2025.

\bibitem{Fedorov2021}
A.~Fedorov, W.~J.~R. Longabaugh, D.~Pot, D.~A. Clunie, S.~Pieper, H.~J. W.~L. Aerts, A.~Homeyer, R.~Lewis, A.~Akbarzadeh, D.~Bontempi, W.~Clifford, M.~D. Herrmann, H.~Höfener, I.~Octaviano, C.~Osborne, S.~Paquette, J.~Petts, D.~Punzo, M.~Reyes, D.~P. Schacherer, M.~Tian, G.~White, E.~Ziegler, I.~Shmulevich, T.~Pihl, U.~Wagner, K.~Farahani, R.~Kikinis, Nci imaging data commons., Cancer research 81 (2021) 4188--4193.
\newblock \href {https://doi.org/10.1158/0008-5472.CAN-21-0950} {\path{doi:10.1158/0008-5472.CAN-21-0950}}.

\bibitem{Gorman2023}
C.~Gorman, D.~Punzo, I.~Octaviano, S.~Pieper, W.~J.~R. Longabaugh, D.~A. Clunie, R.~Kikinis, A.~Y. Fedorov, M.~D. Herrmann, Interoperable slide microscopy viewer and annotation tool for imaging data science and computational pathology., Nature communications 14 (2023) 1572.
\newblock \href {https://doi.org/10.1038/s41467-023-37224-2} {\path{doi:10.1038/s41467-023-37224-2}}.

\bibitem{Clunie2019}
D.~A. Clunie, Dual-personality dicom-tiff for whole slide images: A migration technique for legacy software, Journal of Pathology Informatics 10 (2019) 12.
\newblock \href {https://doi.org/10.4103/jpi.jpi_93_18} {\path{doi:10.4103/jpi.jpi_93_18}}.

\bibitem{appleAV1}
{Apple Inc.}, Apple unveils m3, m3 pro, and m3 max, the most advanced chips for a personal computer, \url{https://nr.apple.com/Di5I4t7da8}, accessed: April 4, 2025 (2023).

\bibitem{nvidiaAV1}
{Nvidia Corporation}, Nvidia video codec sdk v13.0, \url{https://docs.nvidia.com/video-technologies/video-codec-sdk/13.0/read-me/index.html}, accessed: April 4, 2025 (2025).

\bibitem{vulkanAV1}
A.~Abdelkhalek, Khronos releases av1 decode in vulkan video with sdk support for h.264/h.265 encode - khronos blog, \url{https://khr.io/12p}, accessed: April 4, 2025 (2024).

\end{thebibliography}



\end{document}